\documentclass[manuscript]{acmart}
\makeatletter
\renewcommand\@formatdoi[1]{\ignorespaces}
\makeatother
\usepackage{multirow}
\usepackage{graphicx}
\AtBeginDocument{%
  \providecommand\BibTeX{{%
    \normalfont B\kern-0.5em{\scshape i\kern-0.25em b}\kern-0.8em\TeX}}}

\setcopyright{none}
\renewcommand\footnotetextcopyrightpermission[1]{} 
\usepackage{todonotes}
\begin{document}

\title[Minimizing Mindless Mentions]{Minimizing Mindless Mentions: \\
Recommendation with Minimal Necessary User Reviews}

\author{Danny Stax}
\email{d.stax@student.ru.nl}
\affiliation{%
   \institution{Radboud University}
  \country{Netherlands}
 }

\author{Manel Slokom}
\email{m.slokom@tudelft.nl}
\affiliation{%
   \institution{Delft University of Technology}
  \country{Netherlands}
 }

\author{Martha Larson}
\email{martha.larson@ru.nl}
\affiliation{%
   \institution{Radboud University}
  \country{Netherlands}
 }
\renewcommand{\shortauthors}{Stax D., et al.}

\maketitle

\section{Motivating the Minimization of User Reviews}
Recently, researchers have turned their attention to recommender systems that use only minimal necessary data~\cite{larson2017towards,biega2020operationalizing}.
This trend is informed by the idea that recommender systems should use no more user interactions than are needed in order to provide users with useful recommendations.

In this position paper, we make the case for applying the idea of minimal necessary data to recommender systems that use user reviews. 
We argue that the content of individual user reviews should be subject to minimization.
Specifically, reviews used as training data to generate recommendations or reviews used to help users decide on purchases or consumption should be automatically edited to contain only the information that is needed.

The motivation for review minimization comes from large amount of personal information that users include in their reviews. 
Table~\ref{tab:ex} presents an example review from an Amazon data set~\cite{he2016ups}.
The review contains multiple terms that disclose personal information about the person writing the review. 
The terms are listed at the bottom of the table along with the category of personal information that they reveal. 
Note that such information is privacy sensitive or potentially privacy sensitive.
In large enough quantities, the information could possibly deanonymize a reviewer, who writes multiple reviews.

\begin{table}[!h]
\begin{tabular}{|cc|}
\hline
\multicolumn{2}{|c|}{\textbf{Review}}                                                                                                                                                                                                                                                                            \\ \hline
\multicolumn{2}{|c|}{\begin{tabular}[c]{@{}c@{}}Being a 63 year old woman with arthritic hands I think  I did really well in getting this chair put together. (...) \\ I am 5'5" and when I set the chair so my feet rest on the floor, (...) but it is also big enough for my larger husband\end{tabular}} \\ \hline
\multicolumn{1}{|c|}{\textbf{Category}}                                                                                                                              & \textbf{Term}                                                                                                                             \\ \hline
\multicolumn{1}{|c|}{Age}                                                                                                                                & 63 year old                                                                                                                                           \\
\multicolumn{1}{|c|}{Gender}                                                                                                                                      & Woman                                                                                                                                        \\
\multicolumn{1}{|c|}{Medical Information}                                                                                                                            & Arthritic Hands                                                                                                                           \\
\multicolumn{1}{|c|}{Physical Information}                                                                                                                                       & I am 5'5"                                                                                                                                        \\
\multicolumn{1}{|c|}{Marital Status}                                                                                                                                    & Husband                                                                                                                                \\ \hline
\end{tabular}
\caption{Example of potentially sensitive personal information included in a user review.}
\label{tab:ex}
\vspace{-0.25cm}
\end{table}

Note that the main information in the review is that the item (a chair) is easy to assemble and suitable for people of different sizes.
Although, mention of ``arthritic hands'' may be relevant, it is not necessary for the reviewer to state that she is 63 or married in order to communicate the basic message.
We call the unnecessary personal information that users include in their reviews ``mindless mentions" because it is not related to the main message of the review.
Further, users are probably not aware of how personal information accumulates over time.
In the privacy literature, the exploitation of such accumulation is known as an aggregation of information attack~\cite{henriksen2016re}. 
Very active users have a higher chance to have more personal attributes inferred and inference can make use of the combination of multiple information sources~\cite{almishari2012exploring}.



Our position is that researchers should devote research attention to understanding how reviews can be minimized without limiting their usefulness for recommender systems or for user decision making. 
In the rest of this paper, we first present the results of a simple experiment that makes plausible that it is possible to minimize user reviews (even radically) without endangering recommender system performance.
Then, we discuss approaches to intelligent review minimization that we find promising for future development.

\section{The surprising usefulness of minimized reviews}
We evaluate the performance of a simple content-based recommender using full reviews and using reviews that have been radically minimized by removing words. 
Our choice of radical minimization is \emph{word type removal} and is inspired by the example in Table~\ref{tab:ex}, in which we see that personal information is often expressed with nouns and numbers. 
Specifically, we minimize reviews by eliminating these word types: Pronouns, proper nouns, nouns, verbs and numerals. This removes about 70\% of the words in all the reviews. Detecting these word types was done by using a part-of-speech tagger~\cite{honnibal2017spacy}.
Note that we do not eliminate adjectives or adverbs, which we find to be generally less privacy sensitive. We conjecture that adjectives and adverbs carry information about user sentiment toward the item, which is important for recommendation.
We compare \emph{word type removal}, with reviews from which 50\% of the words have been randomly removed (\emph{random removal}) as well as with the original reviews.

The details of our experimental setup are as follows.
Each item is represented by a tf-idf vector calculated on a concatenation of all user reviews for that item, which had been pre-processed with stop-word removal and stemming.
We rank items according to a conventional item-item score~\cite{ekstrand2011collaborative},
\begin{equation}
    \tilde{p}_{u,i} = \sum_{j\in I_u}s(i,j)
    \label{eq:ranking}
\end{equation}
where $\tilde{p}_{u,i}$ is the score for user $u$ on item $i$, $I_u$ are the items rated by $u$, and $s$ is the cosine similarity.
Our data sets are described in Table~\ref{tab:dataset}.
We use a temporal 60/20/20 training/validation/test split. 
Evaluation is carried out by adding the target item to a set of 100 candidate items, and then ranking using Eq. \ref{eq:ranking}, following a one-plus-random strategy~\cite{bellogin2011precision}.
\begin{table*}[h]

\vspace{-0.1cm}
\begin{tabular}{llllllll}
\hline
Data set                         & \# Users & \# Items & Density &\# Reviews & \begin{tabular}[c]{@{}l@{}}\# Words \\ Per Review\end{tabular} & \begin{tabular}[c]{@{}l@{}}\# Words \\ Per User\end{tabular} & \begin{tabular}[c]{@{}l@{}}\# Words \\ Per Item\end{tabular}  \\ \hline
Amazon Office Products          & 4905     & 2420  & 0.449\%    & 53258      & 145.51                                                            & 1579.92                                                        & 145.24                                                          \\
Amazon Health and Personal Care & 38609    & 18534  & 0.048\%   & 346355     & 94.56                                                             & 848.32                                                         & 85.87                                                           \\ \hline
\end{tabular}
\caption{Data set statistics. Columns labeled \# Words give the average number of words before removal. }
\label{tab:dataset}
\end{table*}

\begin{table}[!h]
\centering
\begin{tabular}{crrrrrr}
\hline
\multicolumn{1}{l}{}   \textbf{Data Sets}    & \multicolumn{3}{c}{\textit{Office Products}} & \multicolumn{3}{c}{\textit{Health and Personal Care}} \\ \hline
\textbf{Removal Strategy}  & \textbf{MRR}       & \textbf{Recall} & \textbf{Hit Rate}      & \textbf{MRR}            & \textbf{Recall}  & \textbf{Hit Rate}         \\ \hline
\textit{Unaltered Reviews} & 0.1032             & 0.1230  &  0.2269              & 0.0980                  & 0.2194  &    0.3174                \\ \hline
\textit{Random Removal}    & 0.1034             & 0.1270   &    0.2318           & 0.0989                  & 0.2144  &      0.3125              \\ \hline
\textit{Wordtype Removal}  & 0.1099             & 0.1491   &    0.2599           & 0.1409                  & 0.3012    &      0.4198            \\ \hline
\textit{MostPop}           & 0.0252             & 0.0998  &     0.1645           & 0.1118                  & 0.2686     &      0.3827           \\ \hline
\end{tabular}%
\caption{Performance of a simple TopN (N= 10) content-based recommender trained on user reviews measured in terms of Mean Reciprocal Rank (MRR), Recall and Hit Rate. Removing words from reviews does not have serious consequences. A popularity baseline is included for comparison.}
\label{tab:recsys_res}
\end{table}

Table~\ref{tab:recsys_res} reports our results.
The performance of the original reviews, \emph{random removal}, and \emph{word type removal} is largely comparable. 
In some cases, \emph{word type removal} improves the recommender. 
A further experiment not reported here using only adjectives and adverbs performed slightly worse than \emph{word type removal}, suggesting that these word types are indeed important.
Conditions that remove words from reviews are surprisingly competitive with a popularity baseline, which ranks items by number of reviews. 
Note that Office Products is a particularly challenging data set for our simple content-based recommender and scores that hover around the random baseline of Recall $10/101 = 0.0990$.
For Health and Personal Care and Digital Music (not shown here) our recommender beats the random baseline.


\vspace{-0.2cm}
\section{Vision for Intelligent Review Minimization} 
Our experiment with radical removal demonstrates that it is feasible to develop approaches that minimize reviews but do not hurt recommendation.
However, radical removal has two issues.
First, removing nouns might be problematic for recommenders that use topics or aspects extracted from reviews~\cite{mcauley2013hidden, chen2015recommender}.
Second, user reviews have to be interpretable and readable for other users.
For these reasons, researchers must develop more intelligent review minimization strategies that maintain the important information, rather than simply removing lots of words. 

A simple, yet more intelligent, removal strategy would be to use a dictionary of privacy sensitive words, instead of removing all words of a given word types.
Under such a strategy ``husband'' would be removed, but a word that describes an important aspect of a product, such as ``battery'' would remain.
Discovery of new privacy sensitive words could be automatized using techniques such as word embeddings or training a detector akin to a named entity recognizer (NER), that could identify privacy sensitive words in context.

Research on NER for review minimization could build upon previous work in automatically detecting sensitive information in texts.
For example, ~\cite{baumer2017privacy} discusses detection on online reviews of physicians, which often contain sensitive medical information.
According to the authors, physician review websites are mandated by law to use control mechanisms which automatically identify user statements which harm a persons identity or privacy. 
To combat this risk, the authors applied Named Entity Recognition (NER) together with Natural Language Pattern Detection, to identify and mask the revealing information in these user reviews. 
Similar work~\cite{Sotolar2021Anonymization} has been carried out on Czech social messaging data, where NER was combined with rule based matching to detect mentions of information related to one of four personal attributes: \textit{name}, \textit{ID}, \textit{Location}, and \textit{Contact Information}.

Removal strategies suffer the shortcoming that they probably make the reviews unreadable.
To address this issue, two techniques can be investigated.
First, 
assuming we have a good detection algorithm for privacy sensitive information, which classifies the occurrences into categories as in NER, we can mask the privacy sensitive words with generic words from the same categories.
This leaves us with a review which is still usable for a reader, as well as for a recommender system, while having attained increased privacy for the writer of the review.
Second, researchers can adapt summarization technology to summarize reviews in a way that leaves the information important for users intact and removes potentially sensitive information, without making the reviews unreadable.

We also encourage researchers in the social sciences to investigate why users include personal information in reviews.
The review in Table~\ref{tab:ex} reads like a social media posts, which implies that part of the motivation of the author in writing the review is sharing her personal situation in detail. 
Previous work has suggested that people's knowledge of privacy is not fully in line with their social media sharing behavior~\cite{debatin2009facebook}.
Researching user motivation would reveal which and to which extent mentions can really be considered ``mindless'', as we are assuming and whether it makes sense to try to raise awareness among users of the risks of writing personal information in reviews.


\bibliographystyle{ACM-Reference-Format}
\bibliography{paper}


\begin{thebibliography}{13}


\ifx \showCODEN    \undefined \def \showCODEN     #1{\unskip}     \fi
\ifx \showDOI      \undefined \def \showDOI       #1{#1}\fi
\ifx \showISBNx    \undefined \def \showISBNx     #1{\unskip}     \fi
\ifx \showISBNxiii \undefined \def \showISBNxiii  #1{\unskip}     \fi
\ifx \showISSN     \undefined \def \showISSN      #1{\unskip}     \fi
\ifx \showLCCN     \undefined \def \showLCCN      #1{\unskip}     \fi
\ifx \shownote     \undefined \def \shownote      #1{#1}          \fi
\ifx \showarticletitle \undefined \def \showarticletitle #1{#1}   \fi
\ifx \showURL      \undefined \def \showURL       {\relax}        \fi
\providecommand\bibfield[2]{#2}
\providecommand\bibinfo[2]{#2}
\providecommand\natexlab[1]{#1}
\providecommand\showeprint[2][]{arXiv:#2}

\bibitem[Almishari and Tsudik(2012)]%
        {almishari2012exploring}
\bibfield{author}{\bibinfo{person}{Mishari Almishari} {and}
  \bibinfo{person}{Gene Tsudik}.} \bibinfo{year}{2012}\natexlab{}.
\newblock \showarticletitle{Exploring linkability of user reviews}. In
  \bibinfo{booktitle}{\emph{European Symposium on Research in Computer
  Security}}. Springer, \bibinfo{pages}{307--324}.
\newblock


\bibitem[B{\"a}umer et~al\mbox{.}(2017)]%
        {baumer2017privacy}
\bibfield{author}{\bibinfo{person}{Frederik~S B{\"a}umer},
  \bibinfo{person}{Nicolai Grote}, \bibinfo{person}{Joschka Kersting}, {and}
  \bibinfo{person}{Michaela Geierhos}.} \bibinfo{year}{2017}\natexlab{}.
\newblock \showarticletitle{Privacy matters: detecting nocuous patient data
  exposure in online physician reviews}. In
  \bibinfo{booktitle}{\emph{International Conference on Information and
  Software Technologies}}. Springer, \bibinfo{pages}{77--89}.
\newblock


\bibitem[Bellogin et~al\mbox{.}(2011)]%
        {bellogin2011precision}
\bibfield{author}{\bibinfo{person}{Alejandro Bellogin}, \bibinfo{person}{Pablo
  Castells}, {and} \bibinfo{person}{Ivan Cantador}.}
  \bibinfo{year}{2011}\natexlab{}.
\newblock \showarticletitle{Precision-oriented evaluation of recommender
  systems: an algorithmic comparison}. In \bibinfo{booktitle}{\emph{Proceedings
  of the 5th ACM conference on Recommender systems}}.
  \bibinfo{pages}{333--336}.
\newblock


\bibitem[Biega et~al\mbox{.}(2020)]%
        {biega2020operationalizing}
\bibfield{author}{\bibinfo{person}{Asia~J Biega}, \bibinfo{person}{Peter
  Potash}, \bibinfo{person}{Hal Daum{\'e}}, \bibinfo{person}{Fernando Diaz},
  {and} \bibinfo{person}{Mich{\`e}le Finck}.} \bibinfo{year}{2020}\natexlab{}.
\newblock \showarticletitle{Operationalizing the legal principle of data
  minimization for personalization}. In \bibinfo{booktitle}{\emph{Proceedings
  of the 43rd International ACM SIGIR Conference on Research and Development in
  Information Retrieval}}. \bibinfo{pages}{399--408}.
\newblock


\bibitem[Chen et~al\mbox{.}(2015)]%
        {chen2015recommender}
\bibfield{author}{\bibinfo{person}{Li Chen}, \bibinfo{person}{Guanliang Chen},
  {and} \bibinfo{person}{Feng Wang}.} \bibinfo{year}{2015}\natexlab{}.
\newblock \showarticletitle{Recommender systems based on user reviews: the
  state of the art}.
\newblock \bibinfo{journal}{\emph{User Modeling and User-Adapted Interaction}}
  \bibinfo{volume}{25}, \bibinfo{number}{2} (\bibinfo{year}{2015}),
  \bibinfo{pages}{99--154}.
\newblock


\bibitem[Debatin et~al\mbox{.}(2009)]%
        {debatin2009facebook}
\bibfield{author}{\bibinfo{person}{Bernhard Debatin},
  \bibinfo{person}{Jennette~P Lovejoy}, \bibinfo{person}{Ann-Kathrin Horn},
  {and} \bibinfo{person}{Brittany~N Hughes}.} \bibinfo{year}{2009}\natexlab{}.
\newblock \showarticletitle{Facebook and online privacy: Attitudes, behaviors,
  and unintended consequences}.
\newblock \bibinfo{journal}{\emph{Journal of computer-mediated communication}}
  \bibinfo{volume}{15}, \bibinfo{number}{1} (\bibinfo{year}{2009}),
  \bibinfo{pages}{83--108}.
\newblock


\bibitem[Ekstrand et~al\mbox{.}(2011)]%
        {ekstrand2011collaborative}
\bibfield{author}{\bibinfo{person}{Michael~D Ekstrand}, \bibinfo{person}{John~T
  Riedl}, \bibinfo{person}{Joseph~A Konstan}, {et~al\mbox{.}}}
  \bibinfo{year}{2011}\natexlab{}.
\newblock \showarticletitle{Collaborative filtering recommender systems}.
\newblock \bibinfo{journal}{\emph{Foundations and Trends{\textregistered} in
  Human--Computer Interaction}} \bibinfo{volume}{4}, \bibinfo{number}{2}
  (\bibinfo{year}{2011}), \bibinfo{pages}{81--173}.
\newblock


\bibitem[He and McAuley(2016)]%
        {he2016ups}
\bibfield{author}{\bibinfo{person}{Ruining He} {and} \bibinfo{person}{Julian
  McAuley}.} \bibinfo{year}{2016}\natexlab{}.
\newblock \showarticletitle{Ups and downs: Modeling the visual evolution of
  fashion trends with one-class collaborative filtering}. In
  \bibinfo{booktitle}{\emph{proceedings of the 25th international conference on
  world wide web}}. \bibinfo{pages}{507--517}.
\newblock


\bibitem[Henriksen-Bulmer and Jeary(2016)]%
        {henriksen2016re}
\bibfield{author}{\bibinfo{person}{Jane Henriksen-Bulmer} {and}
  \bibinfo{person}{Sheridan Jeary}.} \bibinfo{year}{2016}\natexlab{}.
\newblock \showarticletitle{Re-identification attacks—A systematic literature
  review}.
\newblock \bibinfo{journal}{\emph{International Journal of Information
  Management}} \bibinfo{volume}{36}, \bibinfo{number}{6}
  (\bibinfo{year}{2016}), \bibinfo{pages}{1184--1192}.
\newblock


\bibitem[Honnibal et~al\mbox{.}(2020)]%
        {honnibal2017spacy}
\bibfield{author}{\bibinfo{person}{Matthew Honnibal}, \bibinfo{person}{Ines
  Montani}, \bibinfo{person}{Sofie Van~Landeghem}, {and}
  \bibinfo{person}{Adriane Boyd}.} \bibinfo{year}{2020}\natexlab{}.
\newblock \bibinfo{booktitle}{\emph{spaCy: Industrial-strength Natural Language
  Processing in Python}}.
\newblock
\newblock
\shownote{\url{https://spacy.io/}, Online; accessed 04-August-2022}.


\bibitem[Larson et~al\mbox{.}(2017)]%
        {larson2017towards}
\bibfield{author}{\bibinfo{person}{Martha Larson}, \bibinfo{person}{Alessandro
  Zito}, \bibinfo{person}{Babak Loni}, {and} \bibinfo{person}{Paolo
  Cremonesi}.} \bibinfo{year}{2017}\natexlab{}.
\newblock \showarticletitle{Towards minimal necessary data: The case for
  analyzing training data requirements of recommender algorithms}. In
  \bibinfo{booktitle}{\emph{FATREC 2017 Workshop on Fairness, Accountability,
  and Transparency in Recommender Systems, in conjunction with the 11th ACM
  Conference on Recommender Systems (RecSys)}}.
\newblock


\bibitem[McAuley and Leskovec(2013)]%
        {mcauley2013hidden}
\bibfield{author}{\bibinfo{person}{Julian McAuley} {and} \bibinfo{person}{Jure
  Leskovec}.} \bibinfo{year}{2013}\natexlab{}.
\newblock \showarticletitle{Hidden factors and hidden topics: understanding
  rating dimensions with review text}. In \bibinfo{booktitle}{\emph{Proceedings
  of the 7th ACM conference on Recommender systems}}.
  \bibinfo{pages}{165--172}.
\newblock


\bibitem[Sotol\'{a}\v{r} et~al\mbox{.}(2021)]%
        {Sotolar2021Anonymization}
\bibfield{author}{\bibinfo{person}{Ond\v{r}ej Sotol\'{a}\v{r}},
  \bibinfo{person}{Jarom\'{\i}r Plh\'{a}k}, {and} \bibinfo{person}{David
  \v{S}mahel}.} \bibinfo{year}{2021}\natexlab{}.
\newblock \showarticletitle{Towards Personal Data Anonymization for Social
  Messaging}. In \bibinfo{booktitle}{\emph{Text, Speech, and Dialogue: 24th
  International Conference, TSD 2021, Olomouc, Czech Republic, September 6–9,
  2021}}. \bibinfo{publisher}{Springer}, \bibinfo{pages}{281–292}.
\newblock
\showISBNx{978-3-030-83526-2}


\end{thebibliography}


\end{document}